\documentclass[12pt]{JHEP3}
\usepackage{amsmath}

\newcommand{\be}[1]{ \begin{equation}\label{#1} }
\newcommand{\ee}{\end{equation}}
\newcommand{\bea}[1]{\begin{eqnarray}\label{#1} }
\newcommand{\eea}{\end{eqnarray}}
\newcommand{\eq}[1]{(\ref{#1})}
\def\ZZZ{{\hskip-3pt\hbox{ Z\kern-1.6mm Z}}}
\def\zzz{{\hskip-3pt\hbox{ z\kern-1mm z}}}

\newcommand{\half}{{1\over 2}}

\def\one{{\hbox{ 1\kern-.8mm l}}}
\def\zero{{\hbox{ 0\kern-1.5mm 0}}}

\title{An AdS$_3$ Dual for Minimal Model CFTs}

\author{
Matthias R.\ Gaberdiel$^{a,b}$ and Rajesh Gopakumar$^{c}$ \\ 
$^a$School of Natural Sciences, \\
$\;$Institute for Advanced Study, \\
$\;$Princeton, NJ 08540, USA \\ \\ 
$^b$Institut f\"ur Theoretische Physik, ETH Zurich, \\
$\;$CH-8093 Z\"urich, Switzerland \\
$\;$\email{gaberdiel@itp.phys.ethz.ch}\\ \\ 
$^c$Harish-Chandra Research Institute, \\
$\;$Chhatnag Road, Jhusi,\\
$\;$Allahabad, India 211019\\
$\;$\email{gopakumr@hri.res.in}}

\abstract{We propose a duality between the 2d ${\cal W}_N$ minimal
models in the large $N$ 't~Hooft limit, and a family of higher spin
theories on AdS$_3$. The 2d CFTs can be described as WZW coset models, and 
include, for $N=2$, the usual Virasoro unitary series. The dual bulk theory 
contains, in addition to
the massless higher spin fields, two complex scalars (of equal mass). The mass
is directly related to the  't~Hooft coupling constant of the dual CFT. 
We give convincing evidence that the spectra of the two theories match precisely
for all values of the 't~Hooft coupling. 
We also show that the RG flows in the 2d CFT  agree exactly with the usual
AdS/CFT prediction of the gravity theory. Our proposal is in many ways analogous  
to the Klebanov-Polyakov conjecture for an 
AdS$_4$ dual for the singlet sector of large $N$ vector models.}

\preprint{HRI/ST/1011}

\begin{document}

\section{Introduction}

Two dimensional conformal field theories are probably the best understood amongst all 
quantum field theories. The local conformal symmetry described by the Virasoro algebra is 
in many cases powerful enough to lead to a complete determination of the operator spectrum, as well as 
to explicit formulae for the correlation functions. These theories thus give concrete instances of 
nontrivial fixed points of the renormalisation group, many of which have a realisation in statistical 
mechanical systems. 

In higher dimensional CFTs, without the luxury of the local Virasoro symmetry, we have had to 
resort to other techniques to learn about nontrivial fixed points. One of the fruitful
approaches has been to consider theories in which one has very many interacting degrees of 
freedom, the so-called large $N$ limit. For example, for vector models in $2+1$ dimensions
with ${\cal O}(N)$ number of fields, one can infer the existence of nontrivial fixed points in the large 
$N$ limit. In fact, in this limit,  the fixed points are perturbatively accessible, and one can compute, 
in a systematic ${1\over N}$ expansion, anomalous dimensions and correlation functions. Thus 
the ${\rm O}(N)$ vector model exhibits the analogue of the Wilson-Fisher fixed point without having 
to resort to methods such as the $\epsilon$ expansion. 

While it has always been surmised that the large $N$ limit is some kind of mean field like description, 
it was not until the advent of the AdS/CFT duality that one could make this idea precise (at least for gauge theories). This duality 
gives a classical description of the leading large $N$ behaviour. The unexpected feature was that this
was in terms of a higher dimensional theory which typically involves gravity in an asymptotically
AdS spacetime. In the case of matrix valued fields with ${\cal O}(N^2)$ degrees of freedom, 
the relevant description is believed to be in terms of a classical  string theory. If one then takes the 
further limit of ultra-strong coupling ($\lambda \gg 1$), the classical 
(super-)string theory reduces to Einstein (super-)gravity. 
This idea has had tremendous success in the last decade or so, and its repercussions are now 
even being felt in domains once far removed from string theory. 

The connection of the large $N$ limit to gravity, however, remains very mysterious, and our 
current understanding is very much tied to the origins of the duality in D-branes and 
string theory (with all its additional baggage of supersymmetry and so on). One would like to 
have examples which are shorn of any unnecessary ingredients, and which give an idea of how this 
connection comes about. Such `distilled' versions of the gauge-gravity duality are also interesting
from the point of view of applications to realistic systems which often do not involve supersymmetry,
for example. Moreover, since an Einstein gravity dual forces us into the regime of very strong coupling 
one would need to move away from this limit to describe systems with a coupling of order one. Generically, 
this would require us to be in a stringy regime with  a large number of operators of finite anomalous 
dimensions. The technical complications of quantising strings in (asymptotically) AdS spacetimes 
prevents us from studying this regime easily. 

An interesting {\it via media} is afforded by the so-called higher spin theories in AdS spacetimes  
\cite{Vasiliev:2003ev}. These are theories containing (generically) an infinite number of massless 
interacting fields with spin $s\geq 2$ (see \cite{Bekaert:2005vh} for an introduction). It has been 
suggested by several people \cite{Sundborg:2000wp,Witten,Mikhailov:2002bp,Sezgin:2002rt} that these 
theories might be relevant for the description of (a sector of) the weak coupling limit of large $N$ gauge 
theories.  However, a striking and concrete conjecture was made in 2002 by 
Klebanov and Polyakov \cite{Klebanov:2002ja} who suggested 
that a particular higher spin theory on AdS$_4$ might be exactly dual to the singlet sector of the 
interacting (as well as the free) ${\rm O}(N)$ vector model in $2+1$ dimensions at large $N$. This is interesting
for several of the reasons discussed above. The ${\rm O}(N)$ model has a close relation to various statistical 
mechanical systems. The interacting fixed point is nontrivial and yet not strongly coupled. Finally, 
it is a concrete duality which goes beyond the Einstein gravity limit and yet does not involve an entire 
stringy spectrum of operators. Recent calculations have  provided non-trivial, interesting evidence 
for this conjecture, see in particular \cite{Giombi:2009wh,Giombi:2010vg,Koch:2010cy}.

The aim of this paper is to propose another duality of this nature. In fact, we 
shall return to the well understood class of 2d CFTs and look for signatures of a 
higher dimensional classical gravity like description in a suitable large $N$ limit. 
This will give rise to a controlled environment in which to study the 
puzzle of the emergence  of a gravity dual.

The large $N$ limit of various field theories in two and higher dimensions has been 
much studied. For some reason, however, this limit does not appear to have been 
much explored in the context of 2d CFTs (see however 
\cite{Bakas:1989xu,Bakas:1990ry,Kiritsis:2006hy}), 
perhaps because they are solvable by other 
means. In this paper we study a family of minimal model CFTs which are given by 
coset WZW models 
\begin{equation}\label{gencos1}
\frac{{\mathfrak su}(N)_k \oplus {\mathfrak su}(N)_1 }{ {\mathfrak su}(N)_{k+1}} \ ,
\end{equation}
where the denominator is the diagonal ${\mathfrak su}(N)$ subalgebra, and the 
subscripts refer to the level of the current algebra. This family of CFTs includes in the 
special case of $N=2$ the usual coset description of the unitary minimal models $(c <1)$ of 
the Virasoro algebra \cite{Goddard:1986ee}. 
Though the generalisation of these theories to arbitrary $N$ has been less studied 
compared to the $N=2$ case, several important facts about them are known. In particular,
the spectrum of  primaries and the fusion rules follow directly from those
of the WZW models, and the characters can in principle be deduced. More interestingly, 
these theories are known \cite{Bais:1987zk} to possess a higher spin ${\cal W}_N$ symmetry 
\cite{Fateev:1987zh,Bilal:1988ze,Bais:1987dc} (for a review see \cite{Bouwknegt:1992wg}),
and the different  minimal models (for finite, fixed  $N$ and different values of $k$) are 
related to one another by an integrable RG flow.
More details about the ${\cal W}_N$ minimal models are explained in Sec.~2.

Here, we will look at these theories in the large $N$ limit. Specifically, we will define a 
't~Hooft limit (see also \cite{Kiritsis:2006hy}) in which we take 
\be{thftlim}
N,k \rightarrow \infty\ ; \qquad  0\leq  \lambda \equiv {N\over k+N} \leq 1\quad \hbox{fixed.} 
\ee 
It is interesting that the limit appears to be well defined and non-trivial.  In particular, these
theories behave like vector models, since their central charge equals 
$c_N(\lambda) \simeq  N(1-\lambda^2)$ and hence scales as $N$. 
The discrete set of CFTs coalesce into a line labelled by the 't~Hooft coupling $\lambda$, where 
$\lambda=0$ behaves like a free theory (of $N$ complex fermions), 
while $\lambda=1$ is some sort of `strong' coupling region.
Notice that the coupling always remains of order one --- an indication of the absence of a dual 
Einstein gravity regime. Furthermore, the spectrum of primaries simplifies remarkably 
in the 't~Hooft limit, in that the dependence of the conformal dimensions on the coupling $\lambda$ 
becomes essentially linear. 
One of the nice features of this model compared to the ${\rm O}(N)$ vector model is the existence of the
additional continuous parameter $\lambda$, which makes it closer to the supersymmetric gauge 
theories in higher dimensions.  The details of the 
't~Hooft limit are explained in Sec.~3.

In Sec.~4 we describe the higher spin theories on AdS$_3$ which are dual to these large $N$ CFTs. 
As was mentioned before, the CFTs have a higher spin ${\cal W}_N$ symmetry, and so it is natural 
that the bulk theory also possesses such a symmetry. In fact, it was recently pointed out in 
\cite{Henneaux:2010xg, Campoleoni:2010zq} that higher spin theories in AdS$_3$ possess, 
at the classical level, an asymptotic symmetry group which is indeed a two dimensional 
${\cal W}$-algebra. This generalises the observation of Brown-Henneaux for asymptotic 
Virasoro symmetries in Einstein gravity on AdS$_3$ \cite{Brown:1986nw}. 
Here we will consider a theory of higher spins, which contains, in addition to a large 
$N$ tower of massless higher spin gauge fields, two complex scalar fields (of equal mass). 
It is known that scalar fields can appear as additional matter fields in these AdS$_3$ theories (precisely in pairs of equal mass). However, 
their mass cannot be arbitrary since it is related to a parameter $\Delta$ of the algebra 
which plays the role analogous to $\alpha^{\prime}$ 
\cite{Prokushkin:1998bq,Prokushkin:1998vn,Vasiliev:1999ba}.\footnote{We thank Misha Vasiliev for this 
remark.} We will see that this parameter (and hence the mass) is indeed 
mapped to the 't~Hooft coupling \eq{thftlim} of the CFT via
\be{massthft}
\Delta =1-\lambda \qquad \Rightarrow \qquad M^2= -(1-\lambda^2)\ .
\ee 

In Sec.~5 we provide support for this conjecture. The first piece of evidence 
consists in matching the spectrum of the CFT with that of the bulk theory. In fact, a 1-loop computation 
for the higher spin gauge fields in AdS$_3$ \cite{Gaberdiel:2010ar} had already revealed a match 
with the vacuum character of the ${\cal W}_N$ algebra. In the specific higher spin theory being 
considered here one has additional scalar fields in the bulk, as well as additional primary fields in the 
CFT.  We find highly non-trivial evidence that the two match for {\it all} values of the  coupling 
$\lambda$. This requires the spectrum of dimensions in the CFT to take a special form which it 
obligingly does, but only in the large $N$ 't~Hooft limit. 

The next piece of evidence consists of relating the behaviour of the CFTs under the RG flow with that in 
AdS$_3$. The lowest non-trivial primary operator ${\cal O}$ has conformal dimension 
$h_-={\bar h}_- =\half(1-\lambda)$ in the 't~Hooft limit. The `double trace' operator 
${\cal O}{\cal O}^\dagger$ is thus relevant, and it is known to be the operator that 
induces the RG flow to the nearby minimal model. Furthermore, one knows that 
it flows in the IR to an irrelevant operator  of the form 
${\cal O^{\prime}} {\cal O}^{\prime\, \dagger}$, where ${\cal O^{\prime}}$ has 
dimensions (in the 't~Hooft limit)  $h_+={\bar h}_+ =\half(1+\lambda)$. Note that 
\be{opdim}
\Delta_+ =(h_+ + {\bar h}_+) =2-\Delta_- =2-(h_- +{\bar h}_-)\ . 
\ee
This precisely corresponds to what we have learnt from AdS/CFT. In fact, one of the bulk complex 
scalar fields $\phi$ that one has to add to the higher spin theory is precisely dual to 
${\cal O}$, while the other, $\phi^{\prime}$, is dual to ${\cal O^{\prime}}$. Though both fields in 
AdS$_3$ have the same mass, there is a choice in how they are quantised 
\cite{Breitenlohner:1982jf, Klebanov:1999tb}. In fact, we have to quantise them in opposite ways 
such that they correspond to the two different dimensions 
$\Delta_{\pm}$ for the operators dual to them.
As was argued on general grounds in \cite{Witten:2001ua} (see also \cite{Berkooz:2002ug}), the 
RG flow takes one from the quantisation corresponding to $\Delta_-$ in the UV to that for
$\Delta_+$ in the IR. In other words, the operator corresponding to $\phi$, namely ${\cal O}$, must 
flow in the IR to the operator corresponding to $\phi^{\prime}$, namely ${\cal O^{\prime}}$.
This is thus in perfect agreement with the CFT result we mentioned earlier.

In Sec. 6 we outline a heuristic way to `derive' this duality. This uses the fact that the 
bulk description of the higher spin fields is a Chern-Simons theory \cite{Achucarro:1987vz,Witten:1988hc}.
One might therefore  imagine the boundary theory to be a WZW theory. In fact, there is a very specific set 
of  boundary conditions associated with requiring the spacetime to be asymptotically AdS$_3$ --- this
is for example clearly explained in \cite{Banados:1998gg}, see also 
\cite{Henneaux:2010xg,Campoleoni:2010zq}. From 
the point of view of the WZW theory, these boundary conditions lead to a  specific gauging (Hamiltonian 
reduction) which goes by the name of (classical) Drinfeld-Sokolov (DS) reduction. The bulk description 
in terms of the higher spin theories is, of course, classical and we do 
not have a quantum definition of the theory. What we propose is that the quantum version of the 
above classical DS reduction would define the quantum theory. This quantum theory
is believed to be equivalent to the above coset model, provided that the levels are suitably
identified \cite{Bilal:1988ze,Bilal:1988jf,Bilal:1988jg,Balog:1990mu,Balog:1990dq}.
In order to apply this line of reasoning to our situation the main open problem is to understand
how to describe the scalar fields in this formulation. This approach should hopefully lead to 
a detailed understanding of the emergence of gravity and higher spin 
diffeomorphisms in AdS$_3$.

Finally, Sec.~7 contains concluding remarks. We have sequestered various details of the 
CFT and higher spin theories into three appendices.

\section{A Family of Minimal Model Conformal Field Theories}\label{coset}

In this section we describe the 2d CFTs which will be the key players on the field theory side. 
Here we outline some  of their important properties. In the next section we will take the 't~Hooft 
large $N$ limit of these models.

\subsection{The Minimal ${\cal W}_N$ Models}

The CFTs we are interested are the so-called ${\cal W}_N$ minimal models. They
are most easily described in terms of a coset  \cite{Bais:1987zk}
\begin{equation}\label{gencos}
\frac{{\mathfrak g}_k \oplus {\mathfrak g}_1 }{ {\mathfrak g}_{k+1}} \ ,
\end{equation}
where, in order to obtain ${\cal W}_N$ we consider ${\mathfrak g}={\mathfrak su}(N)$. The central 
charge of the coset equals 
\be{centch}
c = \dim({\mathfrak g})  \Bigl[ \frac{k}{k+h^\vee} + \frac{1}{1+h^\vee} - \frac{k+1}{k+1+h^\vee}  \Bigr] \ ,
\ee
where $h^\vee$ is the dual Coxeter number of ${\mathfrak g}$. For ${\mathfrak g}={\mathfrak su}(N)$
we have $h^\vee=N$, and the central charge becomes 
\be{centchsp}
c_N(p)  = (N-1) \Bigl[1- \frac{N(N+1)}{p(p+1)}\Bigr] \leq  (N-1) \ , 
\ee
where we have introduced the parameter $p=k+N \geq (N+1)$ that will sometimes be useful. Note that for 
$N=2$ this is just the familiar unitary series of the Virasoro minimal models that can be described by the 
above GKO construction with ${\mathfrak g}={\mathfrak su}(2)$ \cite{Goddard:1986ee}. 

For the smallest value $p=N+1$, ({\it i.e.}\ $k=1$), we have a theory with central charge 
$c=\frac{2(N-1)}{N+2}$ which has an alternative realisation in terms of a theory of 
${\mathbb Z}_N$ parafermions \cite{Fateev:1985mm}. The other extreme case corresponds to 
$p\rightarrow \infty$ (taking $k\rightarrow \infty$ while keeping $N$ finite), where $c = (N-1)$, and
the symmetry algebra is equivalent to the Casimir algebra of the ${\mathfrak su}(N)$ affine algebra 
at level  $k=1$ \cite{Bais:1987dc,Bais:1987zk}. The Casimir algebra consists of all ${\mathfrak su}(N)$
singlets in the affine vacuum representation. Since the affine algebra is at level one, it can be realised in 
terms of $(N-1)$ free bosons.  

\subsection{The Minimal Model Representations}

The actual coset theory does not just involve the vacuum representation of the coset algebra 
(\ref{gencos}). The other states of the theory fall into highest weight representations of the coset
algebra. These are labelled by $(\rho,\mu;\nu)$, where $\rho$ is a highest weight representation 
(hwr) of ${\mathfrak g}_k$, $\mu$ is a hwr of ${\mathfrak g}_1$, and $\nu$ is a hwr of 
${\mathfrak g}_{k+1}$\footnote{It is important to note though that the states in the coset do 
{\it not} transform under any non-trivial representations of 
${\mathfrak su}(N)$.}. Only those combinations are allowed where $\nu$ appears in the decomposition 
of  $(\rho\oplus \mu)$ under the action of ${\mathfrak g}_{k+1}$. The relevant selection rule is simply 
\begin{equation}
\rho+ \mu - \nu   \in  \Lambda_R({\mathfrak g}) \ ,
\end{equation}
where $\Lambda_R({\mathfrak g})$ is the root lattice of ${\mathfrak g}$. In addition, there are field 
identifications: the two triplets 
\begin{equation}\label{fid}
(\rho,\mu;\nu) \cong (A\rho, A\mu; A\nu) \ ,
\end{equation}
define the same highest weight representation of the coset algebra, provided that $A$ is an 
outer automorphism of the affine algebra corresponding to ${\mathfrak g}_{l}$,
with $l=k$, $l=1$ and $l=k+1$, respectively.  For ${\mathfrak g}={\mathfrak su}(N)$, the group 
of outer automorphisms is ${\mathbb Z}_N$, generated by the cyclic rotation of the
affine Dynkin labels, {\it i.e.}\ the map
\begin{equation}\label{afflab}
[\lambda_0;\lambda_1,\ldots ,\lambda_{N-1}] \mapsto
       [\lambda_1;\lambda_2,\ldots ,\lambda_{N-1},\lambda_0] \ ,
\end{equation}
where the first entry is the affine Dynkin label. In this notation, the allowed highest weight representations 
of  ${\mathfrak su}(N)$ at level $k$ are labelled by 
\begin{equation}
P^+_k({\mathfrak su}(N)) = 
\Bigl\{ [\lambda_0;\lambda_1,\ldots ,\lambda_{N-1}]  \ : \ 
\lambda_j \in {\mathbb N}_0 \ , \quad \sum_{j=0}^{N-1} \lambda_j = k \Bigr\} \ .
\end{equation}
Note that the field identification (\ref{fid}) does not have any fixed points since 
${\mathbb Z}_N$ acts transitively on the highest weight representations of 
${\mathfrak su}(N)$ at level $k=1$.  

\subsection{Conformal Weights}

It is easy to see that for {\em any} choice of highest weight representations 
$(\rho;\nu)$, there always exists a unique $\mu\in P^+_1({\mathfrak su}(N))$, such that 
$\rho+\mu-\nu \in \Lambda_R({\mathfrak su}(N))$. Thus we may label the highest weight 
representations of the coset algebra in terms of unconstrained pairs $(\rho;\nu)$. These
pairs are still subject to the field identifications 
\begin{equation}
(\rho;\nu) \cong (A\rho;A\nu) \ .
\end{equation}
The conformal weight of the corresponding highest weight representation equals then
\begin{equation}\label{hco}
h(\rho;\nu) =  \frac{C_N(\rho)}{N+k} + \frac{C_N(\mu)}{N+1} - \frac{C_N(\nu)}{N+k+1} + n \ ,
\end{equation}
where $C_N(\sigma)$ is the eigenvalue of the quadratic Casimir
operator of ${\mathfrak g}={\mathfrak su}(N)$ --- hence the $N$-dependence --- 
in the representation $\sigma$, see appendix~\ref{DSCas} for our normalisation convention. 
Here the representation 
$\mu\in P^+_1({\mathfrak su}(N))$ 
is uniquely determined by the condition that $\rho+\mu-\nu \in \Lambda_R({\mathfrak su}(N))$.
Furthermore, $n$ is a non-negative integer, describing 
the `height' ({\it i.e.}\ the conformal weight above the ground state) at which the 
${\mathfrak g}_{k+1}$ primary $\nu$ appears in the representation $(\rho\oplus \mu)$.
Unfortunately, an explicit formula for $n$ is not available, but it is not difficult to work out $n$
for simple examples. Alternatively, one may use the Drinfeld-Sokolov description of these models
(that is briefly reviewed in appendix~\ref{DS}). In that language  the highest weight representations 
are labelled by $(\Lambda^+,\Lambda^-) \cong (\rho;\nu)$, and the conformal weights equal
\be{hexpsimp}
h(\Lambda^+,\Lambda^-)={1\over 2p(p+1)}\, 
\Bigl( \, \biggl| (p+1)(\Lambda^+ +\hat\rho) -p(\Lambda^- + \hat\rho)\biggr|^2 -\hat\rho^2 \Bigr) \ ,
\ee
where $\hat\rho$ is the Weyl vector of ${\mathfrak su}(N)$.
For $N=2$ (the Virasoro minimal models), (\ref{hexpsimp}) reduces to the familiar
formula 
\begin{equation}
h(r,s) = \frac{(r (p+1) - sp)^2 - 1}{4 p (p+1) } = h(p-r,p+1-s)
\end{equation}
with $1\leq r \leq p-1$, $1\leq s \leq p$. Here we have identified 
$\Lambda^+ = \frac{(r-1)}{\sqrt{2}}$ and $\Lambda^- =  \frac{(s-1)}{\sqrt{2}}$.
\medskip

In the following, the primary where $\nu=[1,0^{N-1}]={\rm f}$ is the fundamental
representation\footnote{Note that the representation of the affine ${\mathfrak su}(N)$ algebra 
has $N$ entries as in (\ref{afflab}). Below we will mostly 
drop the affine Dynkin label, and use a description in terms of the usual
($N-1$) Dynkin labels for representations of  ${\mathfrak su}(N)$.}
with $\rho=[0^{N-1}] = 0$ the trivial representation will play an important role. 
Then (\ref{hco})  gives --- in this case $\mu={\rm f}$ with $n=0$ 
\begin{equation}\label{h0f}
h(0;{\rm f}) = \frac{C_N({\rm f})}{N+1} - \frac{C_N({\rm f})}{N+k+1} = 
\frac{(N-1)}{2N} \, \Bigl( 1 - \frac{N+1}{N+k+1} \Bigr) \ ,
\end{equation}
where we have used that 
$C_N({\rm f}) = \frac{1}{2}(\Lambda_{\rm f},\Lambda_{\rm f} +2 \hat\rho) = \frac{N^2-1}{2N}$. On
the other hand, for the coset representation with $\rho={\rm f}$ and $\nu=0$,  $\mu$ is
the anti-fundamental representation, $\mu=\bar{\rm f}$, and we get (again with $n=0$) 
\begin{equation}\label{hf0}
h({\rm f};0) = \frac{C_N({\rm f})}{N+k} +  \frac{C_N({\rm f})}{N+1}  = 
\frac{(N-1)}{2N} \, \Bigl( 1 + \frac{N+1}{N+k} \Bigr) \ .
\end{equation}
An example with $n=1$ arises for the case where $\rho=0$ and $\nu={\rm adj}$, the adjoint
representation. Then $\mu=0$ but $n=1$, and we obtain 
\begin{equation}\label{h0adj}
h(0;{\rm adj}) = 1 - \frac{C_N({\rm adj})}{N+k+1} = 1 - \frac{N}{N+k+1}  \ ,
\end{equation}
where we have used that $C_N({\rm adj}) =  h^\vee = N$. Finally, the representation with 
$\rho={\rm adj}$ and $\nu=0$ also has $\mu=0$ and $n=1$, and the conformal weight is 
\begin{equation}\label{hadj0}
h({\rm adj};0) = 1 + \frac{C_N({\rm adj})}{N+k} = 1 + \frac{N}{N+k}  \ .
\end{equation}

\subsection{Fusion Rules and Characters}\label{FuCh}

The fusion rules of the coset theory follow directly from the mother and daughter theory. Indeed,
in terms of the triplets $(\rho,\mu;\nu)$ the fusion rules are simply
\begin{equation}
{\cal N}_{(\rho_1,\mu_1;\nu_1)\, (\rho_2,\mu_2;\nu_2)}{}^{(\rho_3,\mu_3;\nu_3)}
  = {\cal N}_{\rho_1 \rho_2}^{(k)\ \rho_3} \,\, {\cal N}_{\mu_1 \mu_2}^{(1)\ \mu_3} \, \,
  {\cal N}_{\nu_1 \nu_2}^{(k+1)\ \nu_3} \ ,
\end{equation}
where the fusion rules on the right-hand side are those of ${\mathfrak g}_k$, 
${\mathfrak g}_1$ and ${\mathfrak g}_{k+1}$, respectively. Note that the fusion rules are 
invariant under the field identification (\ref{fid}). Since the fusion rules of the level one factor
are just a permutation matrix, we can also directly give the fusion rules for the representatives
$(\rho;\nu)$ as
\begin{equation}
{\cal N}_{(\rho_1;\nu_1)\, (\rho_2;\nu_2)}{}^{(\rho_3;\nu_3)}
  = {\cal N}_{\rho_1 \rho_2}^{(k)\ \rho_3} \,\,  {\cal N}_{\nu_1 \nu_2}^{(k+1)\ \nu_3} \ .
\end{equation}
\smallskip

Closed form expressions for the characters of the minimal ${\cal W}_N$ highest weight
representations are known in terms of branching functions, see for example eq.\ (7.51) of
\cite{Bouwknegt:1992wg}. However, these expressions are often difficult to evaluate explicitly.
In the following we shall mainly be interested in the large $k$ limit of these models, in which
case the low lying terms of the characters simplify. In particular, the vacuum character becomes
in this limit
\begin{equation}\label{vacch}
\chi_{(0;0)}(q) = q^{-\frac{c_N}{24}} \, \Bigl( 
\prod_{s=2}^{N} \prod_{n=s}^{\infty} \frac{1}{(1-q)^n}  + {\cal O}(q^{k+1}) \Bigr) \ ,
\end{equation}
since for $k\rightarrow \infty$ the character is that of the Casimir algebra, see
eq.~(7.18) of \cite{Bouwknegt:1992wg}. For finite $k$ the corrections to this formula
are a consequence of the null-vectors of the ${\mathfrak g}_k$ and ${\mathfrak g}_{k+1}$ factors in
(\ref{gencos}). For the case of the vacuum representation with $\rho=\nu=0$, these appear first at 
height $h=k+1$. 

For the other characters there is a similar formula in terms of branching functions of the
affine level one representation to the horizontal (finite-dimensional) Lie algebra. However, as
far as we are aware, no simple explicit formulae for these branching functions are known.\footnote{We
thank Terry Gannon for discussions about this point.} We have
worked out the first few branching rules for some small representations in appendix~\ref{branch},
and from that we can conclude that 
\begin{eqnarray}
\chi_{(0;{\rm f})}(q) & = & q^{h(0;{\rm f})}  \Bigl( 1 +  q + 2 \, q^2 + 4 \, q^3 + \cdots \Bigr) \label{[1000]} \\[4pt]
\chi_{(0;{\rm adj})}(q) & = & q^{h(0;{\rm adj})}  \Bigl( 1 +  2 \, q + 4 \, q^2  + \cdots \Bigr) \label{[1001]} \\[4pt]
\chi_{(0;[0,1,0^{N-3}])}(q) & = &  q^{h(0;[0,1,0^{N-3}])} 
\Bigl(1 + q + 3 \, q^2 + \cdots \Bigr) \label{[0100]} \\[4pt]
\chi_{(0;[2,0^{N-2}])}(q) & = &  q^{h(0;[0,1,0^{N-3}])} 
\Bigl( q + q^2 + \cdots \Bigr)  \nonumber \\
&=& q^{h(0;[2,0^{N-2}])} \Bigl( 1 + q + \cdots \Bigr) \ . \label{[2000]}
\end{eqnarray}
These formulae will play an important role below.

\subsection{The RG Flows}

For fixed (finite) $N$ the models with different values of $k$ (or $p$) are related to one another by an 
RG flow.  This is most familiar for the Virasoro minimal models, for which the perturbing field
in the UV  is the $(1,3)$ field with $h_{1,3} = \frac{p-1}{p+1}$ 
\cite{Zamolodchikov:1987ti}.\footnote{Here, and in the following, we mean by $(1,3)$ the field
whose left- and right-moving conformal dimension is $h=\bar{h}=h(1,3)$.}
In the above conventions this field corresponds to $(\rho;\nu)=(0;{\rm adj})$, 
which has indeed $h(0;{\rm adj})=\frac{p-1}{p+1}$, see (\ref{h0adj}). The RG flow that is induced by this
relevant perturbation connects the $p$-th unitary minimal model in the UV, to the $(p-1)$st in the IR. 
In the IR,  the perturbing $(1,3)$ field of the UV theory has become irrelevant. Indeed, it can be identified
with the $(3,1)$ field of the $(p-1)$'st minimal model \cite{Zamolodchikov:1987ti}. The latter field
has conformal dimension $h_{3,1} = \frac{p+1}{p-1}$ in the $(p-1)$'st minimal model, and hence
can be identified with the  $(\rho;\nu)=({\rm adj};0)$ field in that theory, see (\ref{hadj0}). 

Similarly, the $(1,2)$-field can be identified with $(\rho;\nu)=(0;{\rm f})$. In the IR it flows to
the $(2,1)$-field of the $(p-1)$'st minimal model \cite{Ludwig:1987gs}. The latter field can be identified with 
the $(\rho;\nu)=({\rm f};0)$ field in that theory. Note that this is compatible with the above since we 
have the fusion rules (for $p\geq 4$)
\begin{equation}\label{sroot}
(1,2) \otimes (1,2) = (1,1) \oplus (1,3) \qquad \qquad
(2,1) \otimes (2,1) = (1,1) \oplus (3,1) \ .
\end{equation}
Thus the normal-ordered product of the $(1,2)$ field with itself is the $(1,3)$ field, and similarly for
the $(2,1)$ and $(3,1)$ field. 
\smallskip 

The generalisation to $N>2$ is believed to follow a similar pattern, see for example \cite{Ahn:1990gn}. 
The relevant field $(0;{\rm adj})$ of the $p$'th ${\cal W}_N$ minimal model induces an RG flow, whose 
IR fixed point is the $(p-1)$'st  ${\cal W}_N$ minimal model. In the IR the perturbing field becomes irrelevant, 
and is to be identified with the  $({\rm adj};0)$ field of the $(p-1)$'st model, {\it i.e.}
\begin{equation}
(0;{\rm adj}) _{p} \qquad \xrightarrow{\text{RG-flow by $(0;{\rm adj})$}} \qquad  ({\rm adj};0) _{p-1}\ . 
\end{equation}
The analogue of the $(1,2)$ field for $N>2$ is slightly more subtle. For $N>2$, charge conjugation
of ${\rm SU}(N)$ is non-trivial, and there are therefore two fields that play that role. Indeed, the analogue of
the fusion rules (\ref{sroot}) are now 
\begin{equation}\label{ffbarten}
(0;{\rm f}) \otimes (0;\overline{\rm f})  = (0;0) \oplus (0;{\rm adj}) \qquad 
({\rm f};0) \otimes (\overline{\rm f};0)  = (0;0) \oplus ({\rm adj};0) \  ,
\end{equation}
where $\overline{\rm f}$ denotes the anti-fundamental representation of ${\mathfrak su}(N)$. Note
that the conformal dimension of the $(0;\overline{\rm f})$ field obviously equals that of 
$(0;{\rm f})$, and similarly, $h(\overline{\rm f};0) = h({\rm f};0)$. The analogue of the RG flow
for the $(1,2)$ field is then
\begin{eqnarray}
(0;{\rm f})_{p} \qquad  &  \xrightarrow{\text{RG-flow by $(0;{\rm adj})$}}  & 
\qquad ({\rm f};0)_{p-1} \\
{}(0;\overline{{\rm f}})_{p} \qquad  &  \xrightarrow{\text{RG-flow by $(0;{\rm adj})$}}  & 
 \qquad (\overline{{\rm f}};0)_{p-1} \ .
\end{eqnarray}
As we shall see, these RG flows have a very nice interpretation in the bulk theory, following
the general analysis of \cite{Witten:2001ua}, see also \cite{Klebanov:2002ja}.

\section{The  Large $N$  't Hooft Limit}

With all of these preparations in place, we can now explain the large $N$ limit we shall be considering. 
If we take $N\rightarrow \infty$ for constant $k$, then
it follows from (\ref{centchsp}) that $c_N\simeq 2 k +{\cal O}(N^{-1})$, remembering that $p=k+N$
\cite{Bakas:1989xu,Bakas:1990ry}. 
In this limit, however, many important fields will have vanishing conformal dimension. For example, 
this will be the case for $(0;{\rm f})$ and $(0;{\rm adj})$, see (\ref{h0f}) and (\ref{h0adj}).

It is therefore more interesting to consider the 't Hooft like limit
(see also \cite{Kiritsis:2006hy}), where we take both 
$N,k\rightarrow \infty$, but keep the (renormalised) 't Hooft coupling 
\be{thft}
\lambda=\frac{N}{k+N} <1 
\ee
fixed.  In this limit we get a family of CFTs with an effectively continuous central charge
\be{centchthft}
c_N(\lambda) \simeq N(1-\lambda^2) < N  \ . 
\ee
Note that the central charge scales as $N$. In this sense, these theories behave
like vector like models (whose degrees of freedom scale as $N$), rather than like
gauge theory models (where the number of degrees of freedom scales as $N^2$). 

Note that the `free case', $\lambda=0$, corresponds to first taking $k\rightarrow \infty$, before
taking $N\rightarrow \infty$. At finite $N$, the limit $k\rightarrow \infty$ leads to a  theory with 
$c=N-1$. In the large $N$ limit, we then expect this theory to have a description in terms of a 
singlet sector of $N$ free complex fermions.\footnote{For large $N$, we may ignore the difference between
${\mathfrak su}(N)_1$ and ${\mathfrak u}(N)_1$. The latter theory has a description in terms of 
$N$ complex free fermions.}  This would be closely analogous to the free vector model 
considered  in \cite{Klebanov:2002ja}.  In our context, the `singlet sector' condition arises automatically 
as a consequence of the coset construction, and does not have to be added in by hand.

\smallskip

Next we turn to the conformal weights. It follows from (\ref{hexpsimp}) that they become in this limit
\begin{equation}\label{2.36}
h(\Lambda^+,\Lambda^-) \simeq \frac{1}{2} \bigl( \Lambda^+ - \Lambda^- \bigr)^2 +
\frac{1}{N+k} \, ( \Lambda^+-\Lambda^-,\hat\rho ) \ .
\end{equation}
Note that the second term is typically at least of the same order, since $\hat\rho^2 = \frac{N(N^2-1)}{12}$. 
For example, for the fields discussed above, we find in this 't Hooft limit
\begin{equation}\label{fundim}
h(0;{\rm f}) = \frac{1}{2} ( 1 - \lambda) \ , \qquad
h({\rm f};0) =  \frac{1}{2} ( 1 + \lambda) \ , 
\end{equation}
as well as 
\begin{equation}\label{adjdim}
h(0;{\rm adj}) =  1 - \lambda \ , \qquad
h({\rm adj};0) =  1 + \lambda \ .
\end{equation}
Obviously, this also agrees with the formulae obtained from the coset description,
eqs.~(\ref{h0f})~--~(\ref{hadj0}). 
\smallskip

We should stress that there are ambiguities in how to define the large $N$ limit, and that 
we have implicitly made a choice in the above. 
For example, for $N=2$, there exist at least two
different (natural) $k\rightarrow \infty$ limits of the unitary minimal models that have
been considered in the literature \cite{Runkel:2001ng,Roggenkamp:2003qp}. They
lead to quite different limit theories: the spectrum of \cite{Runkel:2001ng} is 
continuous, and the resulting theory seems to be similar to Liouville theory
(see also \cite{Schomerus:2003vv,Fredenhagen:2004cj}), while the spectrum of
\cite{Roggenkamp:2003qp} is discrete. Both are believed to lead to consistent correlation
functions, and thus both seem to define viable large $k$ limits. 

While these limits have only been analysed for $N=2$, it is not difficult to see how their respective
analogues would differ in our case. In order to explain this, let us consider the representations 
of the form $(R;R)$, whose conformal dimension equals
\begin{equation}
h(R;R) = C_N(R) \Bigl( \frac{1}{N+k} - \frac{1}{N+k+1} \Bigr) = \frac{C_N(R)}{(N+k)(N+k+1)} \ .
\end{equation}
Since the Casimir $C_N(R)$ is of order ${\cal O}(N)$ (for representations with a finite number of boxes 
in the Young tableau, the coefficient is one half times the number of boxes $B(R)$, see (\ref{Casscal})), 
the conformal weight then behaves in the large $N$ limit as 
\begin{equation}\label{contin}
h(R;R) = \frac{B(R)}{2} \times \frac{\lambda^2}{N} \ , 
\end{equation}
where $B(R)$ is an integer. In the 't~Hooft limit, both $N$ and $k$ become large, and hence
representations $R_N$ with an arbitrarily large number of boxes $B(R_N)$ are allowed. There 
are now essentially two possibilities we can consider: we can either define the fields of the limit 
theory to be those associated to a family of representations $R_N$ with fixed $B(R_N)$, and then take 
$N\rightarrow \infty$ --- in this case the conformal weight will approach
$h(R_N;R_N)\rightarrow 0$. Or, we consider fields, where, as we take $N\rightarrow \infty$, we also
take $B(R_N)\rightarrow\infty$, keeping only their ratio fixed. The latter prescription leads to a continuous
spectrum (and is the analogue of the proposal of \cite{Runkel:2001ng}), while the 
former leads to a discrete spectrum as in \cite{Roggenkamp:2003qp}. As will become clear
below, the dual of the bulk gravity theory we are about to discuss corresponds to the second option,
{\it i.e.}\ to a limit theory with a discrete spectrum. Indeed, the fields associated to the gravity dual are
those that appear in finite tensor powers of the fundamental (and anti-fundamental) representation,
and therefore $B(R_N)$ will not grow with $N$. 

We should also note that $h=0$ is not the only limit point; for example, for the representations 
of the form $(R\otimes {\rm f};R)$  the conformal dimension behaves as 
\begin{equation}
h(R\otimes {\rm f};R) \simeq \frac{1}{2} + \frac{B(R)}{2} \times \frac{\lambda^2}{N} \ ,
\end{equation}
{\it etc}. Finally, we note that excitations of order $\frac{1}{N}$ are typically seen in symmetric orbifold
CFTs arising from fractionalised momentum; the above behaviour could therefore be
indicative of some string theory interpretation of our higher spin theory.

\section{The Higher Spin AdS$_3$ Dual}

Now we want to switch gears and describe the dual gravity theory for the above
large $N$ family of 2d CFTs; this will turn out to be a higher spin theory. 

Higher spin field theories in three dimensions are relatively more tractable than their 
higher dimensional counterparts.  
Firstly, the massless higher spin fields
themselves  do not contain any propagating degrees of freedom (see {\it e.g.}\ \cite{Gaberdiel:2010ar}
for a recent discussion). 
Secondly, one can (classically) truncate consistently to a finite number of them 
\cite{Aragone:1983sz}. For instance, one can have theories in which one has massless fields of 
spin $s=2,3, \ldots, N$ only, for any $N \geq 2$. Thirdly, there exists a Chern-Simons formulation 
of the classical action for these theories \cite{Blencowe:1988gj, Bergshoeff:1989ns}. 
For theories with a maximal spin 
$N$, the Chern-Simons gauge group is ${\rm SL}(N, {\mathbb R})\times {\rm SL}(N, {\mathbb R})$ 
(in Lorentzian signature) or ${\rm SL}(N, {\mathbb C})$ in Euclidean signature. Thus the interacting 
theory of the higher spin fields can be expressed relatively compactly compared to the higher 
dimensional cases. 

To be a bit more specific, the higher spin gauge fields can be expressed in terms of generalised vielbein 
and connection variables (generalising the familiar case of gravity) 
\be{vielgen}
e_{\mu}^{a_1\cdots a_{s-1}}\ , \qquad \omega_{\mu}^{a_1\cdots a_{s-1}} \ , 
\ee
where $s$ is the spin of the gauge field. In a theory with maximal spin $N$, all these 
variables for the fields with $s=2,3,\ldots, N$ can be packaged together into two 
${\rm SL}(N,{\mathbb R})$  (or one ${\rm SL}(N,{\mathbb C})$, depending on the signature) 
gauge fields. This reflects the fact that all these fields are part of one single multiplet under 
the higher spin symmetry. The action is then given by
\be{csact}
S=S_{\rm CS}[A]-S_{\rm CS}[\tilde{A}]  \ ,
\ee
where 
\be{usualcs}
S_{\rm CS}[A]={k_{\rm CS}\over4\pi}\int {\rm Tr} (A\wedge dA+{2\over 3}A\wedge A\wedge A) \ .
\ee
The Chern-Simons level $k_{\rm CS}$ (to be distinguished  from the $k$ that appeared in the 
previous section) is related to the AdS radius by the classical relation
\begin{equation}
k_{\rm CS}={\ell\over 4G_N}\ .
\end{equation}

In \cite{Campoleoni:2010zq} (see also \cite{Henneaux:2010xg})
it was argued, using the above Chern-Simons formulation, that the theory with maximal spin 
$N$ has an asymptotic ${\cal W}_N$ symmetry algebra.
Already at the classical level one sees a centrally extended algebra with a central charge
whose value was determined to be the same as the Brown-Henneaux result for Einstein gravity on
AdS$_3$
\begin{equation}
c={3\ell\over 2G_N}\ .
\end{equation}
In  appendix~\ref{HS} we summarise, for completeness, the salient features of the 
frame-like formulation and its relation to the more conventional Fronsdal description in terms of 
symmetric  tensor fields of higher rank. 
\smallskip

So far we have only discussed pure higher spin theories. 
In three dimensions  one can also have, in addition to the higher spin fields,
{\it separate} matter multiplets  (for a survey of these matters see \cite{Vasiliev:1999ba}). 
While in higher dimensions the matter fields always lie in 
the same multiplet as the higher spin fields, in three dimensions
the matter multiplet is distinct and contains {\it only}  scalar and/or fermion fields.  
Moreover, the fields in the matter multiplet can be massive since they are not in the 
same representation as the gauge fields.  The mass is related to a deformation 
parameter\footnote{The deformation parameter is sometimes
denoted by $\nu$ in the literature on higher spin theories \cite{Vasiliev:1999ba}. We suggestively call it
$\Delta$ here since it has exactly the same relation to the mass as the conformal dimension of the 
boundary operator.} of the higher spin algebra as 
\be{masspar}
M^2= \Delta(\Delta-2) \ .
\ee   
Typically the matter multiplet contains four scalars, two with mass (\ref{masspar}), and two 
with $M^2=\Delta(\Delta+2)$. These scalars can additionally transform under a global symmetry group. 
However, it is consistent to truncate this multiplet\footnote{We thank Misha Vasiliev for discussions
about this point.}  to just the two scalars 
of mass (\ref{masspar}), and this is what will be relevant for the following.
The interacting theory of 
these scalars with the higher spin fields was constructed in \cite{Prokushkin:1998bq,Prokushkin:1998vn}.
Finally, we should mention
that for generic $\Delta$, it is no longer possible to truncate the massless fields to a maximal spin. 
Thus once we have added such fields (as we are about to do), we have to take the 
$N\rightarrow \infty$ limit. 

We can now describe the higher spin theory we are interested in. It contains, in addition to 
the higher spin gauge fields, a matter multiplet containing
two complex scalar fields of the same mass (\ref{masspar}). 
We will take 
$M^2$ to lie in the window 
\be{scalmss}
-1 \leq M^2 \leq 0\ .
\ee
As is by now familiar from various AdS/CFT applications  
this implies that there are two alternative 
conformally invariant  quantisations (which we denote by $(\pm)$) of these scalar fields. 
These correspond to the two different roots 
$\Delta_{\pm}$ of (\ref{masspar}) determining the asymptotic fall-off behaviour.  We shall
take one of the scalars, which we call 
$\phi$, in the $(-)$-quantisation and thus corresponding to $\Delta=\Delta_-$. The
other scalar, $\phi^{\prime}$, will be taken in the $(+)$-quantisation corresponding to 
$\Delta=\Delta_+$. We will denote this particular one parameter family of theories by 
${\rm HS}(M^2)$. 

Our proposal can now be stated as follows. The ${\cal W}_N$ minimal model CFT with 
't~Hooft  coupling $\lambda$ is dual, in the large $N$ 't~Hooft limit, to the ${\rm HS}(M^2)$ theory with 
the identification 
\begin{equation}\label{prop1}
\Delta_-=1-\lambda \ , \qquad \Delta_+ = 1 + \lambda \ .
\end{equation}
Note that both scalars have the same mass which is given by 
\begin{equation}\label{cent}
M^2=-(1-\lambda^2)\ .
\end{equation}
Before we begin to discuss this proposal further, let us note that both the CFT and the higher spin theory
have the 
same ${\cal W}_\infty$ symmetry. It then makes sense to identify the central charges; this leads to 
\be{centid}
c_{\rm bulk}={3\ell\over 2G_N}=c_N(\lambda)=N(1-\lambda^2)\ .
\ee
The bulk theory is only well-defined in the large $N$ limit (since we can only add massive
scalar fields in this limit). Note that large $N$ means that $G_N$ is small $\sim {1\over N}$ 
(in units where $\ell=1$); thus the large $N$ limit is indeed the semi-classical limit, where
one can trust the bulk description. For finite $N$, we may 
view the CFT  (in its full ${1\over N}$ expansion) as the quantum definition of the higher 
spin theory. 

In the next section we will present some non-trivial checks of the
proposal at leading order in $N$. We shall also give a heuristic derivation of some
parts of the duality in Sec.~6.

\section{Checks of the Proposal}

In this section we shall subject the above proposal to essentially two consistency checks. 
First we shall explain in quite some detail (see section~5.1) that the spectrum of the two theories 
agrees.  More specifically, we shall study the quantum 1-loop partition function of the higher spin theory,
and see how it reproduces the full CFT spectrum in the 't~Hooft limit. This is quite a detailed 
consistency check, and it probes much of the structure of the CFT. The second consistency
check concerns the RG-flow for which we observe a beautiful matching with the bulk 
analysis (section~5.2). 

\subsection{The Spectrum}

In this section we want to calculate the 1-loop partition function of the higher spin theory
and compare it to the full CFT spectrum. There are basically two parts to this calculation. 
For the higher spin fields, the 1-loop determinant was computed recently in 
\cite{Gaberdiel:2010ar}  using the heat kernel techniques of  \cite{David:2009xg}. For 
$N\rightarrow \infty$ the answer is 
\begin{equation}\label{HSpart}
Z_{\rm HS}^{(1)} = \prod_{s=2}^{\infty} \prod_{n=s}^{\infty}  \frac{1}{|1 - q^{n}|^2} \ .
\end{equation}
The higher spin theory ${\rm HS}(M^2)$ we are interested in also contains two complex scalar
fields, one corresponding to $\Delta = \Delta_+$ and one with $\Delta=\Delta_{-}$, see (\ref{prop1}). 
The 1-loop contribution from each complex 
scalar field is \cite{Giombi:2008vd} (see also \cite{David:2009xg})
\begin{equation}\label{scalchar}
Z_{\rm scalar}^{(1)} = \prod_{l=0,l'=0}^{\infty} \frac{1}{(1 - q^{h+l} \bar{q}^{h+l'})^2}  \ ,
\end{equation}
where $h=\frac{\Delta}{2}$. Thus defining 
\begin{equation}
h_\pm = \frac{1}{2} \bigl( 1 \pm \lambda \bigr) = \half(1 \pm \sqrt{1+M^2}) \ ,
\end{equation}
the total 1-loop partition function is 
\begin{equation}
Z_{\rm total}^{(1)} = \prod_{s=2}^{\infty} \prod_{n=s}^{\infty}  \frac{1}{|1 - q^{n}|^2} \times
 \prod_{l_1=0,l'_1=0}^{\infty} \frac{1}{(1 - q^{h_- +l_1} \bar{q}^{h_- +l'_1})^2} \times
  \prod_{l_2=0,l'_2=0}^{\infty} \frac{1}{(1 - q^{h_++l_2} \bar{q}^{h_++l'_2})^2}  \ .
\end{equation}
Our claim is that this partition function {\em agrees} with the full CFT partition function
of the ${\cal W}_N$ model in the 't~Hooft limit!

We have so far not managed to find an analytic proof of this statement, but we shall give below
what we regard to be highly non-trivial evidence in favour of this claim. Before we begin with
the detailed checks, we should first explain intuitively why this could be true.

The first factor coming from $Z_{\rm HS}^{(1)}$ can be identified with a (generic) vacuum character of
the ${\cal W}_{\infty}$-algebra \cite{Gaberdiel:2010ar}. In our case, the character of the vacuum 
representation of the coset CFT is {\it not} generic since we consider the limit of rational theories
at finite $k$. However, as was explained in Sec.~\ref{FuCh}, see in particular
eq.\ (\ref{vacch}), the null vectors only modify the answer at height $k+1$, and thus this 
modification does not play any role in the 't~Hooft limit. We therefore conclude that the 
contribution from the higher spin gauge fields --- the first factor of $Z_{\rm total}^{(1)}$ --- 
reproduces precisely the vacuum character from the CFT perspective. 

The full CFT has obviously many additional states;  indeed, the coset representations are labelled by 
the pairs $(R_1;R_2)$, and the full spectrum (at finite $k$ and $N$) will include all such sectors. 
However, given the structure of the fusion rules, all states of the CFT can be obtained by taking
successive fusion products of the generating fields
\begin{equation}\label{genfield}
(0; {\rm f}) \ , \qquad (0; \bar{\rm f}) \qquad \hbox{and} \qquad
({\rm f};0) \ , \qquad (\bar{\rm f};0) \ ,
\end{equation}
where ${\rm f}$ and $\bar{\rm f}$ are the fundamental and anti-fundamental representation 
of ${\mathfrak su}(N)$. Note that in the  large $N$ limit, the two sectors corresponding to 
$(0; {\rm f})$ and $(0; \bar{\rm f})$ (and similarly for $({\rm f};0)$ and $(\bar{\rm f};0)$)  
effectively decouple; at finite $N$, the field $(0;\bar{\rm f})$ obviously appears in the $(N-1)$-fold fusion of 
$(0;{\rm f})$ with itself, but in the 't~Hooft limit we have to include both separately.

Now the key observation is that the conformal dimension of the first two fields in (\ref{genfield})
is $h=h_{-}$, while that of the second two fields is $h=h_{+}$, see 
eq.~(\ref{fundim}). This suggests the identification
\begin{equation}\label{iden1}
 \prod_{l_1=0,l'_1=0}^{\infty} \frac{1}{(1 - q^{h_- +l_1} \bar{q}^{h_- +l'_1})^2} \quad\longleftrightarrow\quad
 (0; {\rm f})^{\otimes s_1} \otimes  (0;\bar {\rm f})^{\otimes s_2} \ ,
 \end{equation}
{\it i.e.}\ that the product on the left gives the contributions of the fusion products involving
multiple copies of $ (0; {\rm f})$ and $(0;\bar {\rm f})$. Similarly, the other term should be identified with
\begin{equation}\label{iden2}
 \prod_{l_2=0,l'_2=0}^{\infty} \frac{1}{(1 - q^{h_+ +l_2} \bar{q}^{h_+ +l'_2})^2} \quad \longleftrightarrow\quad
 ({\rm f};0)^{\otimes r_1} \otimes  (\bar {\rm f};0)^{\otimes r_2} \ .
 \end{equation}
 Putting all factors together then accounts for the full spectrum of the CFT. 
 In the following we want to check this proposal in more detail. We shall consider the different pieces 
in turn. 

\subsubsection*{The Fusion Powers of $(0;{\rm f})$}

The simplest consistency check is to consider the square root of (\ref{iden1}), and confirm
that it reproduces the states that appear in the fusion powers of $(0;{\rm f})$, say. (Obviously,
the analysis is identical for the fusion powers of $(0;\bar{\rm f})$.)
Expanding out the first few terms in (\ref{scalchar}) with $h=h_-$ leads to 
\begin{eqnarray}
Z^{(1)} & = & q^{h} \bar{q}^{h} \Bigl( 1 + q + q^2 + q^3 + \cdots \Bigr) \,  
\Bigl( 1 + \bar{q} + \bar{q}^2 + \bar{q}^3 + \cdots \Bigr) \nonumber \\
& & + q^{2h} \bar{q}^{2h}  \Bigl( 1 + q + 2  q^2 + \cdots \Bigr) \,  
\Bigl( 1 + \bar{q} + 2  \bar{q}^2 + \cdots \Bigr) \nonumber \\
& & + q^{2h+1} \bar{q}^{2h+1} \Bigl(1+q + \cdots\Bigr) \, \Bigl(1 + \bar{q} + \cdots \Bigr)  + \cdots \ .
\end{eqnarray}
In order to identify this with ${\cal W}_N$ characters, we also have to multiply the expression with the
1-loop determinant coming from the higher spin fields, $Z_{\rm HS}^{(1)}$ (\ref{HSpart}). Then the 
low-lying terms 
of $Z^{(1)} \cdot Z_{\rm HS}^{(1)}$ look like the sum of three representations with conformal dimensions 
$h_1=h$, $h_2=2h$ and $h_3=2h+1$, whose
characters are 
\begin{eqnarray}
\chi_{h_1}(q) & = &  q^{h} \Bigl( 1 + q + 2 q^2 + 4 q^3 + \cdots \Bigr) \\
\chi_{h_2}(q) & = &  q^{2h} \Bigl( 1 + q + 3 q^2 + \cdots \Bigr) \\
\chi_{h_3}(q) & = &  q^{2h+1} \Bigl( 1 + q + \cdots \Bigr) \ ,
\end{eqnarray}
respectively. Since $h= h_- = \frac{1}{2}(1-\lambda) =  h(0;{\rm f})$, these characters agree then 
precisely, in the 't~Hooft limit, with  the characters for the representation $(0;{\rm f})$, see (\ref{[1000]}), 
the representation $(0;[0,1,0^{N-3}])$, see (\ref{[0100]}), and the representation
$(0;[2,0^{N-2}])$, see (\ref{[2000]}), respectively. 
Here we have used that the conformal dimension of these fields, in the 't~Hooft limit, are 
\begin{equation}
h(0;[0,1,0^{N-3}] ) =  \frac{(N-2)(N+1)}{N} \Bigl( \frac{1}{N+1} - \frac{1}{N+k+1} \Bigr)
\simeq 1 - \lambda = 2 \, h(0;{\rm f})  \ ,
\end{equation}
as well as 
\begin{equation}\label{0,20}
h(0;[2,0^{N-2}]) = \frac{2(N+k)}{N+k+1} \frac{N-1}{N} - \frac{N-1}{N+k+1} \simeq
2 - \lambda = 2 \, h(0;{\rm f}) + 1 \ .
\end{equation}
Note that these two representations are precisely the representations that appear in the fusion
of $(0;{\rm f})$ with itself,
\begin{equation}
(0;{\rm f}) \otimes (0;{\rm f}) = (0;[0,1,0^{N-3}] ) \oplus (0;[2,0^{N-2}] )  \ ,
\end{equation}
in accordance with the fact that the terms that are proportional to second powers
of $h$ correspond to two-particle states in the bulk. 

\subsubsection*{Higher Orders}

We would expect that this pattern continues for higher powers of $q$ and $\bar{q}$. While we have 
not yet attempted to prove this in general, there is one simple consistency check we 
have performed. Since $h$ has a non-trivial $\lambda$-dependence, the above
can only work out if the $\lambda$-dependence is additive under taking
tensor products. It follows from (\ref{2.36}) that the $\lambda$-dependent term is proportional
to $(\Lambda^+-\Lambda^-,\hat\rho)$. For representations that have a finite number of 
boxes in the Young tableau,  the argument in (\ref{Casscal}) then implies that in the large $N$ limit 
\begin{equation}\label{lasca}
\frac{1}{N+k} (\Lambda^+-\Lambda^-,\hat\rho) \ \simeq  \
 \frac{\lambda}{2} \Bigl( B(\Lambda^+) - B(\Lambda^-) \Bigr) \ ,
\end{equation} 
where $B(\Lambda^\pm)$ is the number of boxes in the Young tableau of $\Lambda^\pm$. 
For representations that appear in finite tensor powers of the fundamental, the number of boxes is 
conserved under taking tensor products (for $N$ sufficiently large), and since the fusion rules do not
mix $\Lambda^+$ and $\Lambda^-$ (that contribute with opposite sign), the statement follows.

\subsubsection*{Fusion Products of $(0;{\rm f})$ and  $(0;\bar{\rm f})$}

It is clear that the above analysis works identically for the other factor in (\ref{iden1}),
the one associated to fusion products of $(0;\bar{\rm f})$. However, in order to check
(\ref{iden1}), we also have to verify that the fusion products involving both 
$(0;{\rm f})$ and  $(0;\bar{\rm f})$ work out. The leading `mixed' term arises
by taking terms proportional to $q^{h+l}\bar{q}^{h+l'}$ from both factors in (\ref{iden1}); 
it is easy to see that their total contribution is precisely 
\begin{equation}\label{cross1}
\frac{q^{2h_-} \bar{q}^{2h_-} }{(1-q)^2 (1-\bar{q})^2}  \ .
\end{equation}
Taking into account the ${\cal W}$-descendants, this then implies that the character of 
the corresponding CFT representation should be (in the 't~Hooft limit)
\begin{eqnarray}\label{adjchar}
\chi_{(0;{\rm adj})}(q) & = & q^{1-\lambda} \frac{1}{(1-q)^2} \prod_{s=2}^{N} \prod_{n=s}^{\infty} \frac{1}{(1-q^n)}  
\nonumber \\
& = & q^{1-\lambda} \Bigl(1 + 2 q + 4 q^2 + \cdots\Bigr) \ . \label{adjpre}
\end{eqnarray}
Because these states are single-particle in each factor, they should arise from the tensor product 
(\ref{ffbarten}), and hence transform in the $(0;{\rm adj})$ representation of the coset algebra --- the 
other representation that appears in this fusion product is the identity representation that is already
accounted for by $Z_{\rm HS}^{(1)}$. This works out precisely (to the order to which we have done
the calculation), because (\ref{adjpre}) agrees exactly with 
the character of $(0;{\rm adj})$, see eq.~(\ref{[1001]}). 

\subsubsection*{Fusion Products of $({\rm f};0)$ and  $(\bar{\rm f};0)$}

It is fairly straightforward to see that the analysis works essentially identically for the terms
in (\ref{iden2}). The main difference is that we now have to determine 
the leading behaviour of the characters of the representations $(R;0)$ with $R$ being
in turn $R={\rm f}$, $\bar{\rm f}$, ${\rm adj}$, {\it etc}. It is not difficult to show that the leading 
behaviour of the character of $(R;0)$ is in fact the same as that for $(0;\bar{R})$. For concreteness, 
let us concentrate on the case when $R={\rm f}$. For $({\rm f};0)$, {\it i.e.}\ $\rho={\rm f}$ 
and $\nu=0$, we have $\mu=\bar{f}$. Then the leading behaviour of the character is described 
by the branching function where we count the multiplicities with which the $\bar{\rm f}$-representation 
appears in the level $k=1$ representation based on $\mu=\bar{\rm f}$, since we have to look
at those representations of the level $k=1$ factor that lead to the trivial representation when
tensored with the ground state representation $\rho={\rm f}$. However, this branching function
is precisely what gives the leading part of the $(0;\bar{\rm f})$ character. 
The other cases
work essentially identically. Thus we conclude that  the contribution from the left-hand-side
of (\ref{iden2}) accounts for the tensor products of $({\rm f};0)$ and $(\bar{\rm f};0)$. 

\subsubsection*{Fusion Product of $({\rm f};0)$ and $(0;\bar{\rm f})$}

Finally we also have to look at the terms that involve both 
contributions from (\ref{iden1}) and (\ref{iden2}).  By the same argument as that leading up to 
(\ref{cross1})  it is clear that the leading term of the gravity calculation is 
\begin{equation}
\frac{q^{h_{+} +h_{-}} \bar{q}^{h_{+}+h_{-}} }{(1-q)^2 (1-\bar{q})^2}  \ .
\end{equation}
Since $h_{+}+h_{-}=1$, and taking into account the ${\cal W}$-descendants, this then 
implies that the character of the corresponding CFT representation should be 
(in the 't~Hooft limit)
\begin{eqnarray}
\chi(q) & = & q^{1} \frac{1}{(1-q)^2} \prod_{s=2}^{N} \prod_{n=s}^{\infty} \frac{1}{(1-q^n)}  
\nonumber \\
& = & q^{1} \Bigl(1 + 2\, q + 4\, q^2 + \cdots\Bigr) \ . \label{adjppre}
\end{eqnarray}
Let us first consider the case $({\rm f};0) \otimes (0;\bar{\rm f}) = (\bar{\rm f};{\rm f})$, in which
case we should expect (\ref{adjppre}) to agree with the character of $(\bar{\rm f};{\rm f})$. 
For $\rho=\bar{\rm f}$, $\nu={\rm f}$, the level $k=1$ representation
is $\mu=[0,1,0^{N-3}]$. In this level $k=1$ affine representation we then have to look for those
representations $S$ of the finite-dimesional ${\mathfrak su}(N)$ algebra that have the property that
\begin{equation}\label{cond1}
\bar{\rm f} \otimes S \supset {\rm f} \ .
\end{equation}
Among the representations $S$ that appear in the affine level $k=1$ representation of 
$\mu=[0,1,0^{N-3}]$, the only ones
that satisfy (\ref{cond1}) are 
\begin{equation}
S = [0,1,0^{N-3}]  \qquad \hbox{and} \qquad
S=[2,0^{N-2}] \ .
\end{equation}
Thus the leading behaviour of the character $\chi_{(\bar{\rm f} ,{\rm f})}$ equals the sum
of the branching functions of the level $k=1$ representation $\mu=[0,1,0^{N-3}]$ into 
$[0,1,0^{N-3}]$ and $[2,0^{N-2}]$, respectively. This can be read off from  
(\ref{010b1}) -- (\ref{010b3}), and at least the first  terms we have worked out 
reproduce precisely (\ref{adjppre}). 
Note that the conformal dimension of the primary of the $(\bar{\rm f} ,{\rm f})$ representation is 
\begin{eqnarray}
h(\bar{\rm f} ;{\rm f}) & =  & \frac{C_N(\bar{\rm f})}{N+k} + \frac{C_N([0,1,0^{N-3}])}{N+1} 
- \frac{C_N({\rm f})}{N+k+1}  \nonumber \\
& = & \frac{N^2-1}{2N} \frac{1}{(N+k)(N+k+1)} + \frac{1}{N+1} \frac{(N-2)(N+1)}{N}  \simeq  1 
\end{eqnarray}
in the 't~Hooft limit, thus accounting also correctly for the $q^1$ leading power. 

\subsubsection*{Fusion Product of $({\rm f};0)$ and $(0;{\rm f})$}

For the case where we consider instead the fusion product 
\begin{equation}
({\rm f};0) \otimes (0;{\rm f}) = ({\rm f};{\rm f})
\end{equation}
the gravity calculation is identical. However, now the CFT character is different. For
$({\rm f};{\rm f})$, we have $\mu=0$, and we have to look for the multiplicities with which 
$S=[0^{N-1}]$ and $S=[1,0^{N-3},1]$ appear in the decomposition of the level $k=1$ 
vacuum representation. Actually, by the argument leading to (\ref{adjpre}), the latter contribution 
corresponds precisely to (\ref{adjppre}), and thus we have 
\begin{equation}\label{tri}
\chi_{({\rm f};{\rm f})}(q) = \chi_{(0;0)}(q) + q^{1} \Bigl(1 + 2 q + 4 q^2 + \cdots\Bigr) \ .
\end{equation}
The fact that the limit character decomposes in this manner suggests that the underlying
representation $({\rm f};{\rm f})$ becomes reducible in this limit. Actually, this
phenomenon is familiar from the $k\rightarrow\infty$ limit of the $N=2$ unitary minimal
models, see \cite[Remark 4.1.7]{Roggenkamp:2003qp}. What it means is that in the limit 
$k\rightarrow\infty$, the representation $({\rm f};{\rm f})$  contains a `null-vector'  that
generates the subrepresentation corresponding to the second sum in (\ref{tri}). A
natural way to deal with these additional null-vectors was proposed in  
\cite{Roggenkamp:2003qp} (see also \cite{Graham:2001tg}), where it was referred to as 
`scaling up the additional null vectors'. It amounts to rescaling the states in such a way
that {\em only} the descendants of the null-vectors, {\it i.e.}\ the second sum in (\ref{tri}) survives
in the limit.\footnote{In addition, there will be an overall infinite normalisation factor, reflecting 
the volume divergence of the gravity calculation (that has been dropped in these 1-loop
calculations).}
This is precisely what the gravity calculation also seems to require. We therefore
find again perfect agreement between the CFT and the gravity calculation.

\subsection{The RG Flow}

As mentioned in Sec.~2.5, the minimal models have an RG flow relating two nearby 
theories (labelled by $p$ and $p-1$). The operator responsible for the flow is the least relevant operator 
$(0;{\rm adj})$. In the 't Hooft limit we saw in (\ref{adjdim}) that its dimension becomes $h= 1-\lambda$.
Combining with the the similar operator for the right mover we have a relevant operator for the full CFT. 
In the 't~Hooft limit, the RG flow going from $p$ to $p-1$ changes the 't~Hooft coupling as 
\be{lamchng}
\delta \lambda = {\lambda^2 \over N}\ .
\ee 
Though the 't~Hooft coupling only changes infinitesimally in the large 
$N$ limit, there is nevertheless an order one change in the central charge
\be{centchng}
\delta c = -2\lambda^3\ .
\ee
This indicates that one should be able to see a reflection of this RG flow in the bulk as well. 

As we have seen above, see (\ref{ffbarten}), the field $(0;{\rm adj})$ is actually 
the normal ordered product of the two fields $(0;{\rm f})$ and $(0;\bar{\rm f})$. Writing
${\cal O} = (0;{\rm f})$, and using that $(0;\bar{\rm f})$ is the conjugate of $(0;{\rm f})$, 
the full perturbation is of the form 
\be{cftpert}
S_{\rm pert} =g \int d^2 z\,\, {\cal O}\, {\cal O}^{\dagger} \ ,
\ee
where both ${\cal O}$ and ${\cal O}^{\dagger} $ have conformal dimension
$\half(1-\lambda, 1-\lambda)$ (see (\ref{fundim})). Thus the perturbation is indeed by
a `double trace' operator \cite{Aharony:2001pa}. 
As is familiar from general AdS/CFT considerations, see
in particular \cite{Witten:2001ua}, it corresponds to a flow between two different bulk theories.
The scalar field corresponding to ${\cal O}$ (with dimension 
$\Delta_{-}$) is quantised  
in the $(-)$-quantisation in the UV. Under the RG flow, it flows to a theory with the
$(+)$-quantisation in the IR where it corresponds to an irrelevant operator ${\cal O}^{\prime}$ with 
dimension $\Delta_{+}$.
Here $\Delta_{\pm}$ are the two roots of the equation $M^2=\Delta(\Delta-2)$ for the mass of 
the bulk scalar field. 

This now ties in perfectly with what we know of the corresponding RG flow in the 2d CFT. As 
was explained in Sec.~2.5, the RG flow takes the operator $(0;{\rm adj})$ to $({\rm adj};0)$,
and indeed $(0;{\rm f})$ to $({\rm f};0)$ (as well as $(0;\bar{\rm f})$ to $(\bar{\rm f};0)$). 
Translated into the above language it follows that the operator 
${\cal O}^{\prime}$ can be identified with $({\rm f};0)$ (and similarly for the conjugate field). 
This operator has conformal dimension $\half(1+\lambda,1+\lambda)$, see (\ref{fundim}). 
In particular, we see that it is dual to the scalar field in the $(+)$-quantisation, as
expected. Note that this statement only holds in the 't~Hooft limit, where we have the 
relation $h(0;{\rm f})+h({\rm f};0)=1$ (and similarly $h(\bar{\rm f};0)+h(0;\bar{\rm f})=1$). 

Thus the two complex scalar fields in the bulk with mass $M^2=-(1-\lambda^2)$, where
one is in the $(+)$-quantisation and the other in the $(-)$-quantisation,
fit exactly with what one expects from general consideration of RG flows in AdS/CFT. Furthermore,
the picture ties in perfectly with what is known about the flow in the dual CFT.

\section{Towards a Derivation}

Finally, we want to sketch a possible way in which one can at least heuristically establish the relation
between the bulk theory of  higher spins in AdS$_3$, and the dual coset models studied in Sec.~2 and 3. 
Our starting point will be the Chern-Simons description of the higher spin theory which was 
mentioned in Sec.~4. Let us consider, for definiteness, the Lorentzian signature theory with 
gauge group ${\rm SL}(N, {\mathbb R})\times {\rm SL}(N, {\mathbb R})$. As in Sec.~4 we denote 
the corresponding gauge fields by $A$ and $\tilde{A}$. We will be interested in taking the large 
$N$ limit eventually (to consistently couple with matter) but for now we will take $N$ to be finite for 
simplicity. 

In describing the bulk gravity (or higher spin theories) in a Chern-Simons formulation 
it is absolutely crucial to specify the boundary conditions properly. Since there are no 
propagating modes in the bulk, all the dynamics essentially arises from the boundary 
conditions. For the case of pure gravity on AdS$_3$ (corresponding to $N=2$) this has been 
carefully studied over the years, and there is  a reasonably straightforward generalisation for any 
value of $N$ \cite{Campoleoni:2010zq, Henneaux:2010xg}. We will mainly follow the 
very clear presentation by \cite{Banados:1998gg}, and refer to this paper as well as 
 \cite{Campoleoni:2010zq, Henneaux:2010xg} for more details as well as references to the 
 original literature. 
 
We will work in coordinates where the global AdS$_3$ metric reads as 
\be{adsmet}
ds^2= \ell^2\Bigl(1+{r^2\over \ell^2}\Bigr)\, dt^2 -\Bigl(1+{r^2\over \ell^2}\Bigr)^{-1}dr^2 -r^2d\phi^2 \ .
\ee
The boundary is a cylinder parametrised by $t,\phi$ or more naturally 
$w =t-\phi, \tilde{w}=t+\phi$. 
To have a well-defined variational principle for the Chern-Simons action (\ref{csact}), 
we need to either add a boundary term, or specify suitable boundary conditions for the 
gauge fields. The natural choice of  boundary conditions on the boundary cylinder,
which obviates the need for an additional boundary term is 
(see \cite{Banados:1998gg, Campoleoni:2010zq})
\be{csbdy}
A_{\tilde{w}}=0 \ , \qquad  \tilde{A}_{w} =0 \ .
\ee
In other words, the gauge fields $A, \tilde{A}$ have only left-moving and right-moving 
components at the boundary, respectively. It also suggests 
(using eqs.~(\ref{pot}) and (\ref{cspot})) that the gauge fields are effectively in ${\rm SU}(N)$ at the 
boundary \cite{Banados:1998gg}. 

At this stage one is tempted to view the boundary dynamics as that of two chiral 
WZW theories with gauge group ${\rm SU}(N)$. However,  the above boundary conditions 
are not complete, since they do not guarantee that the geometry is asymptotically AdS$_3$, 
{\it i.e.}\ they do not yet include the analogue of the 
Brown-Henneaux boundary conditions. In the case of pure gravity the relevant boundary condition
for the gauge fields in the Chern-Simons formulation was first worked out in 
\cite{Coussaert:1995zp}. This was recently generalised to any $N$ in 
\cite{Campoleoni:2010zq}. Roughly speaking, the additional condition 
removes all components of the gauge field, except for the lowest-spin components
in the decomposition of the algebra with respect to some principal ${\mathfrak sl}(2,{\mathbb R})$
embedding, see eq.~(54) of \cite{Banados:1998gg} and Sec.~4.2 of \cite{Campoleoni:2010zq}. As 
a consequence the WZW model is gauged, and the resulting theory describes the (classical) 
Drinfeld-Sokolov reduction \cite{Campoleoni:2010zq}. This is what is responsible for reducing 
the affine Kac-Moody  algebra to a ${\cal W}_N$-algebra, and by this route the classical
${\cal W}$-algebra for the asymptotic symmetry generators was established
\cite{Henneaux:2010xg, Campoleoni:2010zq}.

At the quantum level, we propose that the analogous statement should involve the 
{\it quantum} Drinfeld-Sokolov reduction of the affine ${\mathfrak su}(N)$ algebra. The 
quantum mechanical treatment of the DS reduction is more involved (and quite different)
from the classical reduction. In particular, it necessitates the introduction of ghosts to take care of 
the constraints (gauging). A full BRST analysis was carried out in \cite{Balog:1990dq}, and we
summarise some of the results in appendix~B. What is important for us is the 
observation \cite{Bilal:1988ze,Bilal:1988jf,Bilal:1988jg,Balog:1990dq, Bouwknegt:1992wg} 
that the CFT at the quantum level is equivalent precisely to the coset theory 
\begin{equation}\label{gencosrep}
\frac{{\mathfrak su}(N)_k \oplus {\mathfrak su}(N)_1 }{ {\mathfrak su}(N)_{k+1}} \ .
\end{equation}
It is important to note here that the level $k$ of the coset theory is not the same as that of the 
quantum DS reduced theory (and therefore of the original Chern-Simons theory), but rather 
that given in (\ref{DScoid}).

Obviously, the discussion so far only involves the higher spin degrees of freedom. 
Our proposal also suggests that we have to add two massive complex scalar fields (of the same mass)
to this theory in order to identify it with the full dual CFT. It would be very interesting to
understand these scalar fields from the point of view of the Chern-Simons theory. This could
then also open the way to a more conceptual understanding of the duality.

\section{Final Remarks}

In this paper we have made a proposal for a duality between a family of 
higher spin theories on AdS$_3$, and a 't~Hooft like limit of a family of 
2d CFTs. More specifically, we have argued that the 't~Hooft limit of the
${\cal W}_N$ minimal models is dual to the higher spin theory on
AdS$_3$, where one adds to the massless higher spin fields 
two massive complex scalars. Unlike the massless higher spin fields, these massive
scalars actually have propagating degrees of freedom. There is a free parameter on
either side, namely the 't~Hooft coupling $\lambda$ of the 2d CFT, and the mass parameter $M$ 
of the massive scalars, and they are directly related to one another, see eq.\ (\ref{cent}).
We have checked this proposal by matching the spectra in quite some detail. We have 
also shown that the RG flow of the 2d CFT, relating the different minimal models
to one another, agrees very nicely with the gravity description. 

Our proposal is in some sense the natural 3d analogue of the 4d higher spin
conjecture of  Klebanov \& Polyakov  \cite{Klebanov:2002ja}.  Indeed, 
at the free point, $\lambda=0$, the 2d CFT is described by some sort of singlet sector 
of free fermions transforming in the fundamental and anti-fundamental representation, and 
is thus the natural lower dimensional analogue of the ${\rm O}(N)$ vector model of 
Klebanov \& Polyakov. However, unlike their case where the duality is only defined for 
two special CFTs, we actually have a full 1-parameter family of conformal fixed points
for which we can identify the dual higher spin theory. Furthermore, our 2d CFTs are limits 
of consistent ${\cal W}_N$ minimal models, and there is no need for a projection to a singlet 
sector. Indeed, the coset construction seems to take  care of this automatically. 

The family of  theories we consider interpolate between the free theory at $\lambda=0$, and the 
$\lambda=1$ theory. The latter, from the point of view of the bulk theory, corresponds to the 
higher spin algebra with massless scalars, $\Delta=M^2=0$. Indeed, non-zero $\Delta$ plays the 
role of $\alpha^{\prime}$ corrections as already mentioned in the introduction. 
It would be good to see in detail how the interpolation between these two limits via the general
interacting theories takes place. From the field theory point of view it would be nice to 
understand the interacting theories from a Landau-Ginzburg picture as in the case 
of $N=2$.\footnote{We thank Shiraz Minwalla for a stimulating discussion on this topic.}
We note that the RG flow between the ${\cal W}_N$ minimal models is integrable \cite{Ahn:1990gn}. 
Thus, one should be able to study the bulk interpretation of the massive two dimensional field theories 
which are in some sense deformed Gross-Neveu models.  

The ${\cal W}_N$ minimal models that appear on the CFT side are non-supersymmetric 
conformal field theories that are believed to describe the multicritical behaviour of 
${\mathbb Z}_N$ symmetric statistical systems \cite{Fateev:1987zh}. Our proposed
duality could therefore also lead to new insights into such statistical systems. 

\smallskip

There are a number of open problems that deserve further study. First of all, it 
would be good to establish the matching of the spectra to all orders, {\it i.e.}\ to
complete the analysis of Sec.~5.  On the bulk side, it is important to write down the 
interactions involving the massive scalar fields in the Chern-Simons formulation which 
appears not to have been done in the higher spin literature. This would also be necessary in 
order to  flesh out 
the arguments of Sec.~6 and thus understand
the underlying mechanism of the duality. The close connection between the ${\cal W}_\infty$
algebra and the algebra of area-preserving diffeomorphisms of 2-surfaces \cite{Bakas:1989xu} 
could be important in this context. 
One could also compute correlation functions in the CFT (in the planar limit) and compare 
with the bulk computation using, for instance, the techniques of 
\cite{Mikhailov:2002bp, Giombi:2009wh, Giombi:2010vg}. As far as we 
are aware, the correlation functions for these ${\cal W}_N$ models have not yet been 
worked out for general $N$, and it would be important to understand their large $N$ limit. 
It is conceivable that the bulk description might give an 
alternative, less tedious way of determining them (at least, in the planar limit), thus 
making the bulk description useful from a practical point of view in 2d CFTs! 
There are also some natural generalisations one can envisage, in particular,
to cases involving fermions and/or supersymmetry, as well as to cosets of 
${\rm O}(N)$ and ${\rm Sp}(N)$, rather than ${\rm SU}(N)$.

\bigskip

{\bf Acknowledgements:} We thank Justin David, Terry Gannon, Tom Hartman, 
Juan Maldacena, Alex Maloney, Shiraz Minwalla, Ingo Runkel, 
Ashoke Sen, Roberto Volpato and Edward Witten for helpful discussions and correspondences.  
We would like to specially thank Misha Vasiliev for very useful correspondences, and for
giving patient answers to all our questions on higher spin theories. 
M.R.G.\ thanks CalTech for hospitality while this work was begun. His stay at the
IAS is partially supported by The Ambrose Monell Foundation. His work is also supported
in part by the Swiss National Science Foundation. R.G.\ would like to thank 
the organisers of the KITP-C (Beijing) workshop on `AdS/CM Duality and other Approaches'
for hospitality while this work was being completed. 
His work was partly supported by a 
Swarnajayanthi Fellowship of the Dept.\ of Science and Technology, Govt.\ of India,
and as always by the support for basic science by the Indian people.

\section*{Appendix} 
\appendix

\section{Higher Spin Theories on AdS$_3$}\label{HS}

Let us first recall some basic features of massless higher spin theories at the non-interacting level
\cite{Fronsdal:1978rb,Vasiliev:1980as} (see for example 
\cite{Campoleoni:2009je, Bekaert:2005vh}  for  reviews and more references). 

The massless spin $s$ fields in three dimensions are completely symmetric tensors 
$\varphi_{\mu_1\mu_2\cdots \mu_s}$ subject to a double trace constraint
\be{dbtr}
\varphi_{\mu_5\cdots \mu_s \alpha\lambda}{}^{\alpha \lambda} = 0 \ .
\ee
This constraint only makes sense if  $s\geq 4$.  In, addition we have a gauge invariance leading to the 
identification of field configurations
\be{ginv}
\varphi_{\mu_1\mu_2\cdots \mu_s} \sim \varphi_{\mu_1\mu_2\cdots \mu_s}
+\nabla_{(\mu_1}\xi_{\mu_2 \cdots \mu_{s})}\ .
\ee
The gauge parameter $\xi_{\mu_2 \cdots \mu_{s}}$ is a symmetric tensor of rank $(s-1)$ which is, 
in addition, traceless, {\it i.e.}\  $\xi_{\mu_3 \cdots \mu_{s}\lambda}{}^{\lambda}=0$. 
This last constraint only makes sense for $s\geq 3$. 

In higher dimensional AdS spacetimes we need to have an infinite tower of these fields to 
obtain classically consistent interacting theories.  It is a special property of AdS$_3$ \cite{Aragone:1983sz} 
that, for every $N \geq 2$, we can have consistent (again classical) truncations to theories 
which have a spectrum containing a single massless field for each spin $s=2,\ldots, N$. 

While it is possible to write down an action for these theories in terms of the Fronsdal fields given 
above, it is much simpler to recast it in terms of  Chern-Simons gauge fields \cite{Blencowe:1988gj}. 
This generalises the observation of \cite{Achucarro:1987vz,Witten:1988hc} 
for the case of pure gravity, {\it i.e.}\ $N=2$, for 
which the Einstein-Hilbert action can be re-expressed in terms of an 
${\rm SL}(2, {\mathbb R})\times {\rm SL}(2, {\mathbb R})$ 
(or ${\rm SL}(2, {\mathbb C})$ in euclidean signature)  Chern-Simons theory.  
In the case of maximal spin $N$ the higher spin theory has an 
${\rm SL}(N,{\mathbb R})\times {\rm SL}(N,{\mathbb R})$ 
(or ${\rm SL}(N,{\mathbb C})$ in euclidean signature) Chern-Simons description. 
Thus the theory is labelled by 
two parameters, $N$ and the AdS radius $\ell$.

Many of the properties of the Chern-Simons description have been reviewed and studied in detail in
\cite{Campoleoni:2010zq}. We just summarise a few of the central points.
One describes the higher spin fields in terms of generalised vielbein and connection variables
\be{vielgen1}
e_{\mu}^{a_1\cdots a_{s-1}} \qquad  \omega_{\mu}^{a_1\cdots a_{s-1}} \ ,
\ee
and relates these to the Fronsdal fields (about an AdS background) via the relation 
(at the linearised level) 
\be{fronsviel1}
\varphi_{\mu_1\mu_2\cdots \mu_s}= 
{1\over s}\, \bar{e}_{(\mu_1}^{a_1}\cdots\  \bar{e}_{\mu_{s-1}}^{a_{s-1}}
e_{\mu_s) a_1\cdots a_{s-1}}\ .
\ee
Here $\bar{e}_{\mu}^{a}$ is the (usual) vielbein for the background AdS$_3$ metric. 
In addition, to the diffeomorphism invariance, the generalised vielbeins and connections
transform under local `frame rotations' which are parametrised by a gauge parameter  
$\Lambda^{b\, a_1\cdots a_{s-1}}$. 

For the Chern-Simons formulation, one considers the combinations
\be{pot}
j_{\mu}^{a_1\cdots a_{s-1}}=\Bigl(\omega+{e\over \ell}\Bigr)_{\mu}^{a_1\cdots a_{s-1}}, \,\,\,\,\,\,\, 
\tilde{j}_{\mu}^{a_1\cdots a_{s-1}}=\Bigl(\omega - {e\over \ell}\Bigr)_{\mu}^{a_1\cdots a_{s-1}} \ ,
\ee
and then defines the gauge potentials   as 
\bea{cspot}
A&=& (j_\mu^a T_a + \cdots +j_{\mu}^{a_1\cdots a_{N-1}}T_{a_1\cdots a_{N-1}})\, 
dx^\mu \nonumber \\[4pt]
\tilde{A}&=& (\tilde{j}_\mu^a T_a +\cdots + \tilde{j}_{\mu}^{a_1\cdots a_{N-1}}T_{a_1\cdots a_{N-1}})\, 
dx^\mu \ .
\eea
Here the $T_a$ are generators of ${\rm SL}(2,{\mathbb R})$, while 
\be{slngen}
T_{a_1\cdots a_{s-1}} \sim T_{(a_1}\cdots T_{a_{s-1})} \ .
\ee
We can thus view the $A, \tilde{A}$ as ${\rm SL}(N, {\mathbb R})$ gauge fields. 
The action of these higher spin fields is given by (\ref{csact}) together with 
(\ref{usualcs}).

Note that in this description in terms of gauge fields, the 
`physical' Fronsdal fields are actually singlets under the diagonal 
${\rm SL}(N, {\mathbb R})$ gauge group. 
This generalises the well known fact that the 
metric field is a singlet under local Lorentz frame rotations (which are 
the diagonal ${\rm SL}(2, {\mathbb R})$ transformations). However, all observables in the bulk 
higher spin theory should be singlets under the 
${\rm SL}(N,{\mathbb R})\times {\rm SL}(N,{\mathbb R})$ (or 
${\rm SL}(N,{\mathbb C})$) gauge 
transformations.

While (\ref{fronsviel1}) gives the relation between the Fronsdal fields and the frame fields (gauge fields)
at the linearised level, there is a generalisation to the full non-linear theory as well -- see 
Sec.~4.3 of \cite{Campoleoni:2010zq}. It was observed there that the Fronsdal fields 
(for the case of $N=3$) are simply expressed in terms of the Casimir generators of ${\rm SL}(3,{\mathbb R})$. 
This is expected to generalise to the case of arbitrary $N$ \cite{Campoleoni:2010zq}. This also 
fits in with the present proposal in which the vacuum sector of the CFT, which contains the 
Casimir algebra of ${\rm SU}(N)$,  describes the pure higher spin field excitations.

\section{The Drinfeld-Sokolov Description}\label{DS}

In the Drinfeld-Sokolov description of the ${\cal W}_N$ theories one starts with some 
WZW model, and then reduces the theory by imposing suitable constraints, see 
{\it e.g.}\ \cite{Bouwknegt:1992wg} for a review. These
constrained WZW models also give a description of Toda theories \cite{Balog:1990mu} that were
known to be closely related to the ${\cal W}_N$ models \cite{Bilal:1988ze}. In the case of 
interest to us, the WZW model is SU$(N)$ at level $k_{\rm DS}$, and in the quantum version
the resulting theory has central charge
\begin{equation}\label{cDS}
c_N(k_{\rm DS}) = (N-1) \Bigl[ 1 - N(N+1) \frac{(k_{\rm DS}+N-1)^2}{(k_{\rm DS} + N)} \Bigr] \ .
\end{equation}
For large $k_{\rm DS}$ the central charge goes as $c_N(k_{\rm DS}) \simeq -  k_{\rm DS}\, N (N^2-1)$; 
for $N=2$ this reduces to the relation $c_2(k_{\rm DS}) \simeq - 6 k_{\rm DS}$ of \cite{Banados:1998gg}, 
where it is also argued that $k_{\rm DS}$ should be chosen to be negative. In order to relate
the Drinfeld-Sokolov construction to the coset construction described in section~\ref{coset}, 
we have to identify 
\begin{equation}\label{DScoid}
\frac{1}{p} = \frac{1}{k+N} = \frac{1}{k_{\rm DS}+N} - 1 \ .
\end{equation}
Then (\ref{cDS}) agrees with (\ref{centch}). 

In the Drinfeld-Sokolov description  the highest weight representations are labelled by 
$(\Lambda^+,\Lambda^-) \cong (\rho;\nu)$ ---  we are using the notation of 
\cite{Bouwknegt:1992wg} ---  and the conformal weight 
equals (see eq.\ (6.74) and (7.53) of \cite{Bouwknegt:1992wg})
\begin{equation}\label{hexp}
h(\Lambda^+,\Lambda^-)=
{c_N\over 24}-{(N-1)\over 24} +{1\over 2p(p+1)}\Bigl|(p+1)(\Lambda^+ +\hat\rho) 
-p(\Lambda^- +\hat\rho)\Bigr|^2 \ , 
\ee
where the central charge $c_N$ is given by (\ref{cDS}), with the relation between $p$, $k$ and 
and $k_{\rm DS}$ being determined by (\ref{DScoid}). Furthermore, $\hat\rho$ is the Weyl vector of 
${\mathfrak su}(N)$, {\it i.e.}\ one half the sum of all positive roots, whose square equals 
$\hat\rho^2= \frac{N(N^2-1)}{12}$. This then leads to (\ref{hexpsimp}). 

\subsection{Examples and Casimirs}\label{DSCas}

It is not difficult to check that the two formulae (\ref{hco}) and (\ref{hexpsimp}) actually agree in simple
cases. For example, both give $h=0$ for the vacuum representation with $(\rho;\nu)=(0;0)$
(and $n=0$). A more interesting case arises if either $\rho$ (=$\Lambda^+$) or $\nu$ (=$\Lambda^-$) 
vanish. In that case, (\ref{hexpsimp}) becomes 
\be{Lminus}
h(0,\Lambda^-)={1\over 2(p+1)}\, \biggl( p(\Lambda^-)^2-2(\Lambda^-,\hat\rho)\biggr)
\ee
or
\be{Lplus}
h(\Lambda^+,0)={1\over 2p}\, \biggl( (p+1)(\Lambda^+)^2+2(\Lambda^+,\hat\rho)\biggr) \ .
\ee
Specialising further to the case that $\Lambda^{\pm}$ equals the fundamental (f) or adjoint (adj)
representation, we then obtain 
\be{hminusfund}
h(0,{\rm f})={(N-1)\over 2N}\, \Bigl(1-{N+1\over p+1}\Bigr) \ ,  \qquad
h({\rm f},0)= {(N-1)\over 2N}\, \Bigl(1+ {N+1\over p}\Bigr) \ , 
\ee
and 
\be{hminusadj}
h(0,{\rm adj})=1-{N\over p+1}\ , \qquad 
h({\rm adj},0)= 1+{N\over p} \ .
\ee
Here we have used that the inverse Cartan matrix $C^{-1}_{ij} = C^{-1}_{ji}$ 
for ${\mathfrak su}(N)$ equals
\begin{equation}\label{suNid}
C^{-1}_{ij} = \frac{i (N-j)}{N} \quad \hbox{for $(i\leq j)$,}  \qquad \hbox{and} \quad
\sum_{i=1}^{N-1} C^{-1}_{ij} = \frac{j}{2}\, (N-j) \ ,
\end{equation}
from which it follows that 
\begin{equation}
(\Lambda_{\rm f})^2={(N-1)\over N} \ , \quad
(\Lambda_{\rm f}, \hat\rho)={(N-1)\over 2}\ , \quad \hbox{and} \quad 
(\Lambda_{\rm adj})^2=2\ , \quad (\Lambda_{\rm adj}, \hat\rho)=(N-1) \ .
\end{equation}
In our conventions the quadratic Casimir is defined to be 
\begin{eqnarray}\label{Casimir}
C_N(\Lambda) & = &  \frac{1}{2} \Bigl[ (\Lambda,\Lambda) + 2 (\Lambda,\hat\rho) \Bigr] \\
& = &  \sum_{i<j} \Lambda_i \, \Lambda_j \frac{i (N-j)}{N}
+ \frac{1}{2} \sum_{j=1}^{N-1} \Lambda_j^2 \frac{j(N-j)}{N} 
+ \sum_{j=1}^{N-1} \Lambda_j \frac{j}{2} \, (N-j)  \ , \nonumber
\end{eqnarray}
where $\Lambda_j$ are the Dynkin labels of the weight $\Lambda$. For weights
that appear in finite powers of the fundamental representation --- these are the 
weights with a finite number of boxes in the Young tableau ---  the leading term in the large 
$N$ limit is 
\begin{equation}\label{Casscal}
C_N(\Lambda) \simeq \frac{N}{2} \sum_{j=1}^{N-1} j \Lambda_j = 
\frac{N}{2} \,  B(\Lambda) \ , 
\end{equation}
where $B(\Lambda)$ denotes the number of boxes in the Young tableau of $\Lambda$.

\section{Branching Functions}\label{branch}

In order to determine the low-lying terms of the coset characters in the large $k$ limit,
we have to determine the decomposition of level $k=1$ affine representations in terms
of representations of the horizontal (zero-mode) algebra. Since the zero modes commute
with $L_0$, we can do this separately level by level. This is to say, we decompose the affine
level $k=1$ representation $\nu$ in terms of $L_0$ eigenspaces as 
\begin{equation}
{\cal H}_{\nu} = \bigoplus_{n=0}^{\infty} {\cal H}^{(n)}_{\nu} \ ,
\end{equation}
and then decompose each ${\cal H}^{(n)}_{\nu}$ under the action of the zero modes. We have
performed this analysis for the first few values of $n$ and some small representations (assuming 
that $N$ is sufficiently large --- for the following $N\geq 5$ will suffice).\footnote{We thank 
Roberto Volpato for helping us check these identities.} Explicitly we find
\begin{eqnarray}
{\cal H}_{[0^{N-1}]}^{(n=0)} & = &  [0^{N-1}] \label{vb1} \\
{\cal H}_{[0^{N-1}]}^{(n=1)} & = &  [1,0^{N-3},1]  \label{vb2}\\
{\cal H}_{[0^{N-1}]}^{(n=2)} & = &  [0,1,0^{N-5},1,0] \oplus 2 \cdot [1,0^{N-3},1]  \oplus [0^{N-1}] 
\qquad  \label{vb3}\\ 
{\cal H}_{[0^{N-1}]}^{(n=3)} & = & [2,0^{N-4},1,0] \oplus [0,1,0^{N-4},2] \oplus 
2\cdot [0,1,0^{N-5},1,0] \nonumber \\
& & \oplus\  4 \cdot [1,0^{N-3},1]  \oplus 2\cdot [0^{N-1}]  \ ,\label{vb4}
\end{eqnarray}
\begin{eqnarray}
{\cal H}_{\rm f}^{(n=0)} & = &  [1,0^{N-2}] \\
{\cal H}_{\rm f}^{(n=1)} & = &  [0,1,0^{N-4},1]  \oplus  [1,0^{N-2}] \\
{\cal H}_{\rm f}^{(n=2)} & = &  [2,0^{N-3},1] \oplus [0^{N-3},2,0] \oplus 2 \cdot [0,1,0^{N-4},1] \oplus
2 \cdot  [1,0^{N-2}] \qquad  \\ 
{\cal H}_{\rm f}^{(n=3)} & = & [1,1,1,0^{N-4}] \oplus [0^{N-3},1,2] \oplus
 2\cdot [2,0^{N-3},1] \oplus [0^{N-3},2,0]  \nonumber \\
 & & \oplus \ 5 \cdot [0,1,0^{N-4},1] \oplus 4 \cdot  [1,0^{N-2}]  \ ,
\end{eqnarray}
and
\begin{eqnarray}
{\cal H}_{[0,1,0^{N-3}]}^{(n=0)} & =  & [0,1,0^{N-3}]   \label{010b1} \\
{\cal H}_{[0,1,0^{N-3}]}^{(n=1)} & =  & 
[2,0^{N-2}] \oplus  [0^{N-3},1,1] \oplus [0,1,0^{N-3}]   \label{010b2} \\
{\cal H}_{[0,1,0^{N-3}]}^{(n=2)} & =  & 
[1,1,0^{N-4},1] \oplus [2,0^{N-2}] \oplus  2 \cdot [0^{N-3},1,1]
\oplus 3\cdot [0,1,0^{N-3}]  \ . \label{010b3}
\end{eqnarray}

\bibliographystyle{JHEP}

\end{document}